\documentclass[prd,showpacs,nofootinbib,preprintnumbers,
preprint,
floatfix]{revtex4}

\usepackage{amsmath}
\usepackage{amsfonts}
\usepackage{color}
\usepackage{dsfont}
\usepackage{appendix}

\begin{document}
\title{Generation of rotating solutions in Einstein-scalar gravity}

\author{Igor Bogush}\email{igbogush@gmail.com}   
 \author{Dmitri Gal'tsov} \email{galtsov@phys.msu.ru}
\affiliation{Faculty of Physics, Moscow State University, 119899, Moscow, Russia}

\begin{abstract}
 
We revisit the problem of the rotating generalization of the Fisher-Janis-Newman-Winicour solution of the minimal Einstein-scalar theory proving that the two previously proposed solutions do not satisfy the equations of motion. We also derive several new spinning solutions with the scalar charge, which are endowed with oblate deformation.  
\end{abstract}

\maketitle

\section{Introduction}\label{sec:intro}
Einstein's general theory of relativity with a minimally coupled massless scalar field (MES) has recently attracted new interest in view of the various dualities that connect this theory with non-minimal scalar-tensor theories such as Horndeski and DHOST.
 \cite{Galtsov:2018xuc,BenAchour:2019fdf}.
This interest gives new life to Fisher's famous solution of MES \cite{Fisher:1948yn}, repeatedly rediscovered in the past by many authors \cite{Bergmann:1957zza,Penney:1968zz,Janis:1968zz,Penney:1969xk,Deser:1969nv,Wyman:1981bd,Virbhadra:1997ie,Bhadra:2001fx}  in particular,  Janis, Newman and Winicour \cite{Janis:1968zz} and often abbreviated as FJNW now. For a more recent discussion of the Fisher solution, see \cite{Abdolrahimi:2009dc}. This solution is singular on the potential horizon, but it may be non-singular in the dual frame of some non-minimal theory \cite{Faraoni:2015paa,Galtsov:2018xuc,Domenech:2019syf}. The frame transformation can be considered as a generation method for finding exact solutions of various non-minimal theories, that probe their physical properties. Therefore, it is interesting to find new physically interesting solutions of the minimal Einstein-scalar theory and, above all, the  Fisher spinning solution. This turned out to be a non-trivial task.  
 
An earlier attempt to introduce the angular momentum into FJNW was made in \cite{Agnese:1985xj} using the Janis-Newman (JN) algorithm \cite{Newman:1965tw} (for a detailed discussion of this method, see \cite{Erbin:2016lzq}). Due to the simplicity of the   \cite{Agnese:1985xj} solution obtained, it was repeatedly  applied in the astrophysical context, see, e.g. \cite{Gyulchev:2008ff,Kovacs:2010xm,Liu:2018bfx,Jusufi:2018gnz}. It is worth noting that the JN method was originally proposed simply as a formal trick leading to Kerr's solution.  Although JN was later tested in various other theories \cite{Erbin:2016lzq}, no rigorous mathematical proof was given in the general case, especially in scalar-tensor theories. Explicit
checking \cite{Pirogov:2013wia} of fulfillment  of a part of Einstein's equations for the solution \cite{Agnese:1985xj} led to the negative result (see also \cite{Hansen:2013owa}). However, since this solution is still used in applications \cite{Liu:2018bfx,Jusufi:2018gnz}, we revise the problem of its validity here, confirming the result of \cite{Pirogov:2013wia}.
Other spinning solutions of the minimal Einstein-scalar theory were found recently \cite{Chauvineau:2018zjy}, one of which  is asymptotically flat. It has a Kerr-like metric, and tends toward the non-spherical Penney solution in the static limit \cite{Penney:1968zz}, so it cannot be considered as a rotating FJNW solution.

Similarly, in the framework of the Brans-Dicke theory (BD), one can construct a rotating charged solution (and more general with the Newman-Unti-Tamburino (NUT) parameter) \cite{Tiwari:1976vc,Kim:1998hc,Kirezli:2015wda} using the Kinnersley form of BD field equations. This solution does not reproduce the FJNW solution in the Einstein frame. A rotating version of the BD analogue of the FJNW solution was built in \cite{Krori:1982} again using the JN trick. We test this solution here.

Hidden symmetries of the static MES equation were observed long ago
\cite{Buchdahl:1959nk, Abdolrahimi:2009dc, Janis:1970kn}, here we give them the modern sigma-model interpretation along the lines of  a more general $\sigma$-model construction for the Einstein-Maxwell-dilaton model (EMD)  \cite{Galtsov:1995mb}. Various mathematically correct generation methods for the minimal Einstein-scalar and the Einstein-Maxwell-scalar system have been proposed in the past based on the dimensional reduction to three and two dimensions. We will briefly review them here and expand to include the Cl\'ement transform \cite{Clement:1997tx} (CT), which generates the Kerr solution from Schwarzschild using the Maxwell field at the intermediate stage of derivation. Generalizing this to include a minimal scalar field, we get some new rotating generalizations of FJNW.  
It turned out that for the successful application of CT to generate rotation in the presence of a scalar field, it is necessary to combine FJNW with the Zipoy-Voorhees (ZV) \cite{Zipoy:66,Voorhees:70} solution, which is similar to FJNW in a spheroidal coordinate system.

FJNW solution is the starting point for other generalizations: for arbitrary dimensions \cite{Xanthopoulos:1989kb}, the Einstein-Maxwell theory    \cite{Penney:1969xk}, the EMD theory \cite{Garfinkle:1990qj}.  FJNW solution has a singular horizon, which only becomes regular in the Schwarzschild limit for a zero scalar charge. It was shown that such objects with singular horizons cannot appear in vacuum  \cite{Winicour:1968zz}, therefore   singular horizons are a feature of theories with  scalar field.  
 
Another method for finding a rotating solution with a nontrivial scalar field we use here was proposed by Eri\c{s} and G\"urses \cite{Eris:1976xj}  (EG-transformation). According to them, stationary axisymmetric EMS solutions obey the equations of motion, which can be divided into purely electrovacuum part and additional terms in metric functions. These terms appear due to the presence of a scalar field that satisfies its equation of motion. Applying this technique to a rotating axisymmetric vacuum solution, we can obtain its generalization with a non-trivial scalar field.

In this paper we define the MES action as
\begin{equation}\label{eq:intro.st_action}
    \mathcal{S} = \int d^4x \sqrt{-g} \left(
        R - 2 (\partial_\mu \phi)(\partial^\mu \phi)
    \right),
\end{equation}
which corresponds to the equations of motion
\begin{subequations} \label{eq:intro.eom}
\begin{align}
    & \label{eq:intro.eom_R}
    R_{\mu\nu}=2(\partial_\mu \phi)(\partial_\nu \phi),
    \\
    & \label{eq:intro.eom_phi}
    \nabla_\mu \nabla^\mu \phi = 0.
\end{align}
\end{subequations}
In the Section \ref{sec:fzv} we combine the ZV and FJNW solutions using the $\sigma$-model technique. In section \ref{sec:rotating_fjnw-zv}  Cl\'ement's transformations will be applied to the solution obtained in the Section \ref{sec:fzv}. As a result, we will get the rotating generalization of the ZV-FJNW solution with the oblateness parameter as a function of mass and  scalar charge. In Section \ref{sec:rotating_fjnw} we will discuss the relation between the FJNW rotating solution and Tomimatsu-Sato (TS) \cite{Tomimatsu:73} and give an example of a rotating FJNW  with a phantom scalar field without oblateness. We will also show that such solutions allow for complex coordinate transformations leading to real scalar field with a different physical meaning.

\setcounter{equation}{0}
\section{Static reincarnations} \label{sec:fzv}
Stationary sector of the model (\ref{eq:intro.st_action}) admits a three-dimensional $\sigma$-model representation,
assuming an ansatz for stationary metrics
\begin{equation} \label{eq:sigma_ansatz}
    ds^2 = -f (dt - \omega_i dx^i)^2 + f^{-1} h_{ij}dx^i dx^j,
\end{equation}
where the function $f$, the one-form $\omega_i$ and the 3-metric $h_{ij}$ are functions of space coordinates $x^i$, $i=1,\,2,\,3$. Indices of 3-metric are supposed to be lowered and raised with 3-metric $h_{ij}$ and 3-inverse metric $h^{ij}$. The 1-form $\omega_i dx^i$ can be expressed in terms of twist potential $\chi$
\begin{equation}
    \partial_i \chi = - \frac{f^2}{\sqrt{h}} h_{ij} \epsilon^{jkl} \partial_k \omega_l,
\end{equation}
where $\epsilon^{jkl}=\pm1$, entering then in a set of three-dimensional scalar potentials $\Phi^A = \{\psi, \chi, \phi\}$ with $\psi=\frac{1}{2}\ln f$ in the action
\begin{equation} \label{eq:sigma_action}
    \mathcal{S} = \int d^3x\, \sqrt{h} h^{ij} \left( R_{ij}^{(3)} - \mathcal{G}_{AB} \partial_i \Phi^A \partial_j \Phi^B \right),
\end{equation}
and the target space metric $\mathcal{G}_{AB}$ given by
\begin{equation} \label{eq:sigma_metric}
    \mathcal{G}_{AB} d\Phi^A d\Phi^B = 2(d\phi^2 + d\psi^2) + \frac{1}{2}e^{-4\psi} d\chi^2,
\end{equation}
where $R_{ij}^{(3)}$ is the Ricci tensor constructed with the metric $h_{ij}$. Note that
MES theory can be considered as a truncation of EMD with trivial electromagnetic field  \cite{Galtsov:1995mb}.

The target-space metric $\mathcal{G}_{AB}$ admits three gauge isometries 
\begin{subequations}
    \begin{align}
        \label{eq:sigma_gauge_1}
        &I:& \phi \to \phi + \lambda_\phi& &\\
        \label{eq:sigma_gauge_2}
        &II:& \chi \to \chi + \lambda_\chi& &\\
        \label{eq:sigma_gauge_3}
        &III:& \psi \to \psi + \lambda_\psi&, \qquad \chi \to e^{2\lambda_\psi} \chi&
    \end{align}
\end{subequations}
with  constant  $\lambda_\phi,\,\lambda_\chi,\,\lambda_\psi$, and a non-trivial Ehlers transformation  \cite{Ehlers:1961zza}
\begin{equation}\label{eq:sigma_ehlers}
    \frac{1}{z} \to \frac{1}{z'} = \frac{1}{z} + i \lambda_E, \qquad
    z = f + i \chi
\end{equation}
with parameter $\lambda_E$.

In the static truncation $\chi=0$
one also has the $SO(2)$-rotational symmetry in the plane $(\psi,\phi)$:
\begin{equation} \label{eq:sigma_so2_transformation}
    \begin{pmatrix}
    \psi\\
    \phi
    \end{pmatrix}
    \to
    \begin{pmatrix}
    \psi'\\
    \phi'
    \end{pmatrix}
    =
    \begin{pmatrix}
    & \cos\beta & -\sin\beta\\
    & \sin\beta & \cos\beta
    \end{pmatrix}
    \begin{pmatrix}
    \psi\\
    \phi
    \end{pmatrix}
\end{equation}
parametrized by the angle $\beta$. Transformation (\ref{eq:sigma_so2_transformation}) is the $\sigma$-model equivalent of the transformation found by Buchdahl   \cite{Buchdahl:1959nk}, which were also rediscovered in \cite{Abdolrahimi:2009dc}.  

\paragraph{Generation of FJNW}

~

Any vacuum solution of general relativity satisfies equations of motion of MES (\ref{eq:intro.eom}) with constant scalar field. Considering the Schwarzschild solution as a seed   and applying the transformation (\ref{eq:sigma_so2_transformation}) with  $\cos\beta = S$, one can recover the FJNW solution in the form (\ref{eq:sigma_ansatz}) with
\begin{subequations} \label{eq:fjnw_spherical}
\begin{equation}
    h_{ij}dx^idx^j = dr^2 + r^2 F \left(d\theta^2 + \sin^2\theta d\varphi^2\right),
\end{equation}
\begin{equation}
    \psi = \frac{S}{2} \ln F,\qquad
    \phi = \phi_\infty - \frac{\Sigma S}{2M} \ln F,\qquad
    \omega_i = 0,\qquad
    S = \frac{M}{\sqrt{M^2 + \Sigma^2}},
\end{equation}
\end{subequations}
where $M$, $\Sigma$ are ADM mass and scalar charge, the function $F$ is
\begin{equation}\label{eq:F_spherical}
    F(r) = 1-\frac{2M}{rS}.
\end{equation}
Setting $\Sigma \to 0$ brings us back to the Schwarzschild solution. 

For further purposes of this paper, it is convenient to use the prolate spheroidal coordinates $x$ and $y$ defined as
\begin{equation}\label{eq:FZV.xy_to_rtheta}
    x=\cfrac{r}{k} - \tilde{k},\qquad
    y=\cos\theta,
\end{equation}
where $k$ and $\tilde{k}$ are constants, chosen so that $g_{\varphi\varphi} = k^2(x^2-1)(1-y^2)$.
The function $F$ and the 3-metric of the solution (\ref{eq:fjnw_spherical}) in prolate spheroidal coordinates with $k = M/S$, $\tilde{k}=1$ read 
\begin{equation}\label{eq:fjnw_spheroidal}
    h_{ij}dx^idx^j = k^2 \left(
        dx^2 + \frac{x^2-1}{1-y^2} dy^2 + (x^2-1)(1-y^2) d\varphi^2
    \right),
\end{equation}
\begin{equation}\label{eq:F_spheroidal}
    F(x) = \frac{x-1}{x+1}.
\end{equation}
This definition   of $F$ will be used in further calculations.

\paragraph{Generation of ZV  with scalar charge}

~

ZV solution in the form (\ref{eq:sigma_ansatz}) reads
\begin{subequations}
\begin{equation}\label{eq:zv_potentials}
    \psi = \frac{\delta}{2} \ln F,\qquad
    \phi = 0,\qquad
    \omega_i = 0,\qquad
\end{equation}
\begin{equation} \label{eq:zv_3_metric}
    h_{ij} dx^i dx^j = k^2\left(
        H_{ZV}(x, y) \left( dx^2 + \frac{x^2-1}{1-y^2} dy^2 \right) + (x^2-1)(1-y^2) d\varphi^2
    \right),
\end{equation}
\begin{equation}\label{eq:zv_H}
    H_{ZV}(x, y) = \left(\frac{x^2-1}{x^2-y^2}\right)^{\delta^2 - 1},
\end{equation}
\end{subequations}
where $k = M/\delta$. One can present the gravitational potential $\psi$ of FJNW solution (\ref{eq:fjnw_spheroidal}) in the same form as ZV (\ref{eq:zv_potentials}) up to an interchange of constants $S$ and $\delta$. This suggests that FJNW and ZV can be naturally combined using the transformations (\ref{eq:sigma_so2_transformation}). Applying the $SO(2)$-transformation to the solution (\ref{eq:zv_potentials}) leads to the ZV metric with a scalar charge which we will denote FZV. It has the potentials
\begin{subequations} \label{eq:fzv_potentials}
    \begin{equation}
        \psi = \frac{S\delta}{2} \ln F,
    \end{equation}
    \begin{equation}
        \phi = \phi_\infty - \frac{\Sigma S \delta}{2M} \ln F,
    \end{equation}
\end{subequations}
the same 3-metric (\ref{eq:zv_3_metric}), and the constant $k = M/S\delta$, giving the ADM mass $M$. The only difference between ZV and FZV solutions is the replacement of the parameters $\delta \to S\delta$ in the gravitational potential $\psi$ and the constant $k$. This modification of the solution will have use for angular momentum generation in the following section.

\paragraph{ Generation of NUT}

~

For  completeness, we add the NUT parameter to the solution (\ref{eq:fzv_potentials}). To do that, we sequentially apply the Ehlers transformation  (\ref{eq:sigma_ehlers}) and the gauge transformation (\ref{eq:sigma_gauge_3}) to ensure $g_{tt}\to-1$ for $r\to\infty$:
\begin{subequations} \label{eq:nut_potentials}
    \begin{align}
        &
        \psi = \frac{1}{2}\ln\frac{(1+\lambda^2)F^{S\delta}}{1 + \lambda^2 F^{2S\delta}},
        \\&
        \omega_i dx^i = 2 N y d\varphi,
        \\&
        \phi = \phi_\infty - \frac{\Sigma}{2k} \ln{F},
    \end{align}
\end{subequations}
where $\lambda$ is the parameter of the Ehlers transformation (\ref{eq:sigma_ehlers}). The 3-metric of the solution with the NUT-parameter has the form (\ref{eq:zv_3_metric}). The ADM mass $M$, the scalar charge $\Sigma$ and the NUT parameter $N$ are
    \begin{equation}
        M = k S\delta \frac{\lambda^2-1}{\lambda^2+1},\qquad
        \Sigma^2 = k^2 \delta^2(1-S^2),\qquad
        N = \frac{2 \delta  k \lambda  S}{\lambda ^2+1},
    \end{equation}
which can be resolved in the following form
\begin{equation}
    k =\frac{\left(\lambda ^2+1\right)^2 M^2-\left(\lambda ^2-1\right)^2 \Sigma ^2}{\delta  \left(\lambda ^2-1\right) \sqrt{\left(\lambda ^2-1\right)^2 \Sigma ^2+\left(\lambda ^2+1\right)^2 M^2}},\qquad
    S=\frac{\left(\lambda ^2+1\right) M}{\sqrt{\left(\lambda ^2-1\right)^2 \Sigma ^2+\left(\lambda ^2+1\right)^2 M^2}}.
\end{equation}

The solution (\ref{eq:nut_potentials}) with $\Sigma = 0$ represents the vacuum ZV solution with NUT parameter, which was given in \cite{Stelea:2006zz}. For $\delta = 1$ we obtain the FJNW solution with NUT found in \cite{Kobyalko:2017zz}.
    
\paragraph{Singularities}

~

The solution (\ref{eq:nut_potentials}) is the most general of all obtained before. The scalar curvature $R$ of this solution can be found from the equation of motion (\ref{eq:intro.eom_R}) as simple as
\begin{equation}
    R = 2 g^{xx} (\partial_x \phi)^2 =
    \frac{2 \Sigma^2}{k^4}
    \frac{(1+\lambda^2)}{1 + \lambda^2 F^{2S\delta}}
    \left(x-1\right)^{S\delta-1-\delta^2}
    \left(x+1\right)^{-S\delta-1-\delta^2}
    \left(x^2-y^2\right)^{\delta^2-1}.
\end{equation}
One can see that the parameter $\lambda$ does not influence the divergence which depends on $S$ and $\delta$  in the exponent of $(x-1)$ and $(x^2-y^2)$. For $y\neq \pm 1$ the metric  is singular for $S\delta - 1 -\delta^2 < 0$. For $S^2 < 1$ this condition is always satisfied. For $y=\pm 1$ the condition is $S\delta < 2$. If $S\delta \geq 2$ the ``horizon'' is not singular.

\paragraph{Chazy-Curzon limit}

~

ZV solution admits the limit $\delta\to\infty$ resulting in the Chazy-Curzon solution \cite{Chazy:1924,Curzon:1925}. Solutions (\ref{eq:fzv_potentials}) and (\ref{eq:nut_potentials}) have the same limiting form. The 3-metric, the scalar field and the function $F^{S\delta}$ will be
\begin{equation} \label{eq:сс_3_metric}
    h_{ij} dx^i dx^j =
        \exp\left\{-\frac{M^2\sin^2\theta}{S^2r^2}\right\} \left( dr^2 + r^2 d\theta^2 \right) +
        r^2 \sin^2\theta d\varphi^2,
\end{equation}
\begin{equation}
    F^{S\delta} \to e^{-2M/r},\qquad
    \phi\to\phi_\infty +\Sigma/r.
\end{equation}

\setcounter{equation}{0}
\section{Cl\'ement transformation} \label{sec:rotating_fjnw-zv}

To generate a rotating solution, here we will apply Cl\'ement's generating technique designed for the Einstein-Maxwell theory \cite{Clement:1997tx}. Since the Maxwell field plays an important role in the procedure, we have to extend the model to  
\begin{equation}\label{eq:RFZV.emd_action}
    \mathcal{S} = \frac{1}{16\pi} \int d^4x \sqrt{-g} \left[
        R - 2(\partial \phi)^2 + F^2
    \right],
\end{equation}
where $F=dA$ is the electromagnetic 2-form. It was shown in \cite{Clement:1997tx} that application of CT to the ZV solution with some oblateness parameter $\delta$ does not lead at the end to any vacuum metric. Still one can hope to be able to do this applying CT to the combined FZV metric with the constant $\delta \to S\delta$. We expect to get rotating generalization of FZV metric imposing the constraint $S\delta=1$. 

First, we generalize the sigma-model to include the Maxwell field introducing the electric and magnetic potentials $v$, $u$  via
\begin{equation}\label{eq:RFZV.em_ansatz}
    F_{i0} = \frac{1}{\sqrt{2}} \partial_i v, \qquad
    F^{ij} = \frac{f}{\sqrt{2h}}\epsilon^{ijk}\partial_k u.
\end{equation}
Other components of electromagnetic tensor in terms of (\ref{eq:RFZV.em_ansatz}) read 
\begin{equation}
    F^{i0} = F^{ij}\omega_{j} - h^{ij}F_{j0}, \qquad
    F_{ij} = f^{-2}h_{ik}h_{jl}F^{kl} + F_{0i}\omega_{j} - F_{0j}\omega_{i},
\end{equation}
where $h^{ij}$ is a 3-inverse metric tensor of $h_{ij}$. We also modify the equations for the twist potential $\chi$ as
\begin{equation}\label{eq:RFZV.chi}
    \partial_i \chi = -f^2 h^{-1/2}h_{ij}\epsilon^{jkl}\partial_k \omega_l + u\partial_i v - v \partial_i u,
\end{equation}
This representation  in terms of scalar potentials $f, \chi, u, v, \phi$ was derived for EMD  in \cite{Galtsov:1995mb}, generalizing the result of \cite{Neugebauer:69}. In our case we have to put $\alpha=0$ obtaining: 
\begin{equation}\label{eq:RFZV.sigma_ems}
    dl^2_{EMS} = \frac{1}{2f^2}\left(df^2 + (d\chi + v du - u dv)^2 \right) - \frac{1}{f}\left( du^2 + dv^2 \right) + 2 d\phi^2.
\end{equation}
Here the scalar field is decoupled from  other potentials, therefore  all symmetries of the Einstein-Maxwell model are preserved. Following \cite{Clement:1997tx} we pass to the  complex Ernst $(\mathcal{E}, \psi)$ and Kinnersley $(U, V, W)$ potentials
\begin{equation}\label{eq:RFZV.ernst}
    \mathcal{E} = f + i \chi - \bar{\psi}\psi 
    =\frac{U - W}{U + W},\quad
    \psi = \frac{v + i u}{\sqrt{2}}
    =\frac{V}{U + W},
\end{equation}
with one of the Kinnersley potentials being redundant. The target space of the $\sigma$-model  (\ref{eq:RFZV.sigma_ems}) possesses the $SU(2,1)$ isometry group, which acts as  on the complex vector space $(U, V, W)$ leaving the norm $\overline{U}U + \overline{V}V - \overline{W}W$ invariant. 
For reader's convenience,  we briefly recall the CT transformation, which is a triple  $\mathcal{R}= \Pi^{-1} R \Pi$ with the target space discrete map
\begin{equation} \label{eq:RFZV.pi}
    \Pi:\; U \leftrightarrow V,
\end{equation}
followed by the coordinate transformation
\begin{equation} \label{eq:RFZV.r}
    R:\; \varphi \to \varphi + \Omega t
\end{equation}
and another target space map.
Both the target-space and the coordinate transformations do not change the scalar field $\phi$ which  depends on $r$ or $x$ only.

Starting with the static vacuum seed solution  $\mathcal{E}_1 \in \mathbb{R}$, $\psi_1 = 0$ we can take
\begin{equation}
    V_1 = 0,\quad
    U_1 = - 1 - \mathcal{E}_1,\quad
    W_1 = - 1 + \mathcal{E}_1,
\end{equation}
where indices number the steps of the procedure. After the first $\Pi$-transformation $U \leftrightarrow V$, the Ernst potentials become
\begin{equation}
    \mathcal{E}_2 = -1,\quad
    \psi_2  = \frac{1 + \mathcal{E}_1}{1 - \mathcal{E}_1}
\end{equation}
and the new functions $f,\chi,u,v$ read  from (\ref{eq:RFZV.ernst}) as
\begin{align}\label{eq:RFZV.potentials_2}
    &
    f_2 = \frac{ 4 \mathcal{E}_1 }{ (\mathcal{E}_1-1)^2 },\quad
    \chi_2 = 0,\quad
    \omega_2 = 0,\quad
    v_2 = \sqrt{2}\frac{ 1 + \mathcal{E}_1 }{ 1 - \mathcal{E}_1 },\quad
    u_2 = 0.
\end{align}    
The corresponding spacetime is not asymptotically flat.
Next, perform  the global coordinate transformation (\ref{eq:RFZV.r}) to a uniformly rotating frame $\varphi = \tilde{\varphi} + \Omega t$.
Acting with (\ref{eq:RFZV.r}) on the metric in the Weyl-Papapetrou parametrization
\begin{equation}
    ds^2 = -f (dt - \omega d\varphi)^2 - f^{-1} \left(\gamma_{mn} dx^m dx^n + \rho^2 d\varphi^2\right)
\end{equation}
  one obtains  
\begin{align*}
    & f' = f [ 1 - 2\Omega\omega+\Omega^2(\omega^2 - f^{-2}\rho^2)],\quad 
    & \omega' = \frac{
        \omega - \Omega(\omega^2-f^{-2}\rho^2)
    }{
         1 - 2\Omega\omega+\Omega^2(\omega^2 - f^{-2}\rho^2)
    }, \\
    &\gamma_{mn}' = \frac{f'}{f}\gamma_{mn} ,&
    \rho'=\rho, \\
    & \partial_m v' = (1 - \Omega\omega)\partial_m v - \Omega f^{-1} \rho \tilde{\partial}_m u,\quad
    & \partial_m u' = (1 - \Omega\omega)\partial_m u + \Omega f^{-1} \rho \tilde{\partial}_m v,
\end{align*}
where $\tilde{\partial}_m = \gamma^{-1/2}\gamma_{mn}\epsilon^{np}\partial_p$. Applying these transformations to (\ref{eq:RFZV.potentials_2}), the transformed functions are simplified to
\begin{equation*}
    f_3 = f_2 ( 1 - w^2 ),\qquad 
    \omega_3 = \frac{ \Omega^{-1}w^2 }{ 1 - w^2 },\qquad
    {\gamma_3}_{mn} = ( 1 - w^2 ) {\gamma_2}_{mn},\qquad
    \rho_3=\rho_2,
\end{equation*}
\begin{equation*}
    v_3 = v_2,\qquad
    \partial_m u_3 = w \tilde{\partial}_m v_2,\qquad
    w = \Omega \rho / f_2.
\end{equation*}

The Ernst potentials are then rescaled by a constant $\mathcal{E}\to p^2 \mathcal{E}$, $\psi \to p \psi$, which corresponds to the solution invariance with respect to the following transformations: $t \to p t$, $f \to p^{-2} f$, $\omega \to p \omega$, $h_{ij} \to p^{-2} h_{ij}$, $u\to p^{-1} u$, $v \to p^{-1} v$. The need for this transformation will be revealed further. Applying all the above transformations to the solution (\ref{eq:fzv_potentials}) and  putting $\mathcal{E}_1 = f_0^{S\delta}$, one  obtains
\begin{align*}
    \mathcal{E}_3 = &
    p^2\left[- 1
    - \left(k y \Omega  S\delta\right)^2
    -\frac{
        k^2 \Omega^2 \left(x^2-1\right) \left(1-y^2\right) (f_0^{S\delta}-1)^2
    }{
        4 f_0^{S\delta}
    }
    + 2 i k y \Omega  \left(
        S\delta \frac{f_0^{S\delta} + 1}{f_0^{S\delta} - 1} + x
    \right)
    \right],
    \\\nonumber
    \psi_3 =  & 
    p \left(
        \frac{1 + f_0^{S\delta}}{1 - f_0^{S\delta}}
        + i k y S\delta \Omega
    \right).
\end{align*}

The last transformation $\Pi^{-1}$ leads to the final solution with the Ernst potentials $\mathcal{E}_4$, $\psi_4$ in the form
\begin{equation*}
    \mathcal{E}_4 = \frac{2\psi_3 + \mathcal{E}_3 - 1}{2\psi_3 - \mathcal{E}_3 + 1},\qquad
    \psi_4 = \frac{1 + \mathcal{E}_3}{2\psi_3 - \mathcal{E}_3 + 1}.
\end{equation*}
To get the final solution with zero electromagnetic field, it is necessary to find such parameters, that set $\psi_4$ equal to zero, which can be achieved with $\mathcal{E}_3=-1$. It is possible only for $S\delta=1$ and $p=\left(1+k^2\Omega^2\right)^{-1/2}$, bringing us to the final expression for the Ernst potentials
\begin{equation}
    \mathcal{E}_4 = \frac{
        p x + i q y - 1
    }{
        p x + i q y + 1
    },
    \quad
    \psi_4 = 0,
\end{equation}
where $q=k\Omega p$ is a constant ($p^2 + q^2 = 1$). Using the transformation from prolate to spherical coordinates in the form $x\to r p / k - \tilde{k}$ and $y \to \cos\theta$ with the redefined constants
\begin{equation*}
    \tilde{k}=M/\sqrt{M^2-a^2},\qquad
    \Omega=aM/(M^2-a^2)^{3/2},\qquad
    k=(M^2-a^2)/M
\end{equation*} and properly rescaling the scalar charge $\Sigma$, the solution will have  the Kerr-like form
\begin{align}\label{eq:rfzv_solution}
    &
    f(r,\theta) = \frac{\Delta - a^2 \sin^2\theta}{r^2 + a^2 \cos^2\theta},
    \qquad
    \omega(r,\theta) = 
    - \frac{
        2 a M r \sin^2\theta
    }{
         \Delta - a^2 \sin^2\theta
    },
    \\\nonumber &
    h_{ij}dx^i dx^j = H(r,\theta) \left( dr^2 + \Delta d\theta^2 \right)
    + \Delta\sin^2\theta d\varphi^2,
    \\\nonumber &
    \Delta(r) = (r-M)^2 - b^2
\end{align}
with the following scalar field $\phi$ and the function $H$:
\begin{equation}\label{eq:rfzv_solution_phi}
    \phi(r) = \phi_\infty + \frac{\Sigma }{2 b} \log \frac{r-M+b}{r-M-b},\qquad
    H(r,\theta) =
    \frac{ \Delta - a^2 \sin^2\theta }{\Delta}\left(
        1 + \frac{ b^2 }{ \Delta }\sin^2\theta
    \right)^{-\Sigma^2/b^2},
\end{equation}
where $b=\sqrt{M^2-a^2}$. Naturally, for $\Sigma=0$ the solution coincides with Kerr solution without scalar field. One can also guess the generalization with the NUT parameter by taking Kerr-NUT solution and putting $\phi$ and $H$ from (\ref{eq:rfzv_solution}), (\ref{eq:rfzv_solution_phi}) with constant $b = \sqrt{M^2 + N^2 - a^2}$. Further in this section we will consider the solution without the NUT charge.

The Ricci scalar following from the equations of motion (\ref{eq:intro.eom_R}) reads
\begin{equation}
    R=2(\partial_r \phi)^2 g^{rr} 
    =
    \frac{2\Sigma^2}{\Delta(r^2 + a^2 \cos^2\theta)} \left(
        1 + \frac{ b^2 }{ \Delta } \sin^2\theta
    \right)^{\Sigma^2/b^2}.
\end{equation}
The curvature scalar $R$ diverges on the ``horizon'' $\Delta = 0,\, b^2 > 0$.
The solution also possesses ring singularities at the equatorial plane: $r=0$ inherited from the Kerr solution; and $r=M$ for $b^2 < 0$.

Recently some new rotating solutions were given by Chauvineau \cite{Chauvineau:2018zjy}. One of them is asymptotically flat, generalizing Penney's solution \cite{Penney:1968zz}. It also has the form (\ref{eq:rfzv_solution}) up to the following redefinition of the scalar field $\phi$ and the function $H$:
\begin{subequations}\label{eq:rfzv_chauvineau}
    \begin{align}
    &\label{eq:rfzv_chauvineau_phi}
    \phi_N = \frac{\Lambda } { \sqrt{\Delta + (M^2 - a^2)\cos^2\theta }},
    \\ & \label{eq:rfzv_chauvineau_h}
    H_N = \frac{ \Delta - a^2 \sin^2\theta }{\Delta} \exp{\left\{
        - \frac{\Lambda^2 \Delta \sin^2\theta } { (\Delta + (M^2 - a^2)\cos^2\theta)^2 }
    \right\}},
    \end{align}
\end{subequations}
where $\Delta = r^2-2Mr+a^2$ is the same as for (\ref{eq:rfzv_solution}), the subscript $N$ stands for ``Newtonian'', following  \cite{Chauvineau:2018zjy},  but we redefined the original definition of constant $\Lambda$ which was $\Lambda\sqrt{M^2-a^2}$ in \cite{Chauvineau:2018zjy}. The scalar curvature for this solution diverges on the equator of the ``horizon''.

Passing to an ``extremal'' limit of solution (\ref{eq:rfzv_solution}) $M^2-a^2 \to 0$, in which the function $H$ and the scalar field are non-trivial:
\begin{subequations}\label{eq:rfzv_extr_solution}
    \begin{align}
        &\label{eq:rfzv_extr_solution_h}
            H =
            \frac{ \Delta - a^2 \sin^2\theta }{\Delta}\exp{\left\{-\frac{\Sigma^2 \sin^2\theta}{\Delta}\right\}},
        \\&\label{eq:rfzv_extr_solution_phi}
            \phi = \phi_\infty + \frac{\Sigma}{\sqrt{\Delta}},
    \end{align}
\end{subequations}
we observe that our solution and that of (\ref{eq:rfzv_chauvineau}) coincide.

\setcounter{equation}{0}
\section{Eris-Gurses transformation}\label{sec:rotating_fjnw}

Further simplification of equations of motion can be achieved with the Weyl-Papapetrou ansatz, assuming the axial symmetry:
\begin{equation}
    ds^2 = -\exp{\left(2\psi\right)}(dt - \omega d\varphi)^2 + \exp{\left(-2\psi\right)}\left[
        \exp{\left(2\gamma\right)}(d\rho^2 + dz^2) + \rho^2 d\varphi^2
    \right],
\end{equation}
where $\psi$, $\omega$ and $\gamma$ are functions of coordinates $\rho=k\sqrt{(x^2-1)(1-y^2)}$ and $z=kxy$. The equations of motion read 
\begin{subequations}\label{eq:eom_weyl}
\begin{equation} \label{eq:eom_weyl_1}
    \Delta \psi = -\frac{1}{2}e^{4\psi}\rho^{-2}(\nabla \omega)^2,
\end{equation}
\begin{equation}\label{eq:eom_weyl_2}
    \nabla\left(e^{4\psi} \rho^{-2} \nabla \omega \right) = 0,
\end{equation}
\begin{equation}\label{eq:eom_weyl_3}
    \gamma_{,\rho} =\rho\left(
        \psi_{,\rho}^2 - \psi_{,z}^2 + \phi_{,\rho}^2 - \phi_{,z}^2 - \frac{1}{4}e^{4\psi}\rho^{-2} \left(\omega_{,\rho}^2 - \omega_{,z}^2\right)
    \right),
\end{equation}
\begin{equation}\label{eq:eom_weyl_4}
    \gamma_{,z} =2\rho\left(
        \psi_{,\rho}\psi_{,z} + \phi_{,\rho}\phi_{,z} - \frac{1}{4}e^{4\psi}\rho^{-2} \left(\omega_{,\rho}\omega_{,z}\right)
    \right),
\end{equation}
\begin{equation}\label{eq:eom_weyl_5}
    \Delta \phi = 0,
\end{equation}
\end{subequations}
where the vector operators act the same way as in the flat space cylindrical coordinates $(\rho, z, \varphi)$. Eris and Gurses \cite{Eris:1976xj} suggested to split the equations into the vacuum and scalar parts. To do that, we have to present the function $\gamma$ as a sum of two terms $\gamma = \gamma^\psi + \gamma^\phi$. Then one can formulate the following theorem. If the functions $\psi$, $\omega$ and $\gamma=\gamma^\psi$ fulfil vacuum equations, then the functions $\psi$, $\omega$ and $\gamma=\gamma^\psi + \gamma^\phi$ satisfy the equations (\ref{eq:eom_weyl}) if the scalar field fulfils the equation (\ref{eq:eom_weyl_5}) and
\begin{equation}\label{eq:eom_weyl_gamma_phi}
    \gamma^\phi_{,\rho} =\rho\left( \phi_{,\rho}^2 - \phi_{,z}^2 \right),\qquad
    \gamma^\phi_{,z} =2\rho \phi_{,\rho}\phi_{,z}.
\end{equation}
The integrability condition of (\ref{eq:eom_weyl_gamma_phi}) is given by (\ref{eq:eom_weyl_5}). 

This theorem allows to generate a solution with non-trivial scalar field from a vacuum solution. We will abbreviate this as EG-transformation. The EG duality can be applied also in the inverse direction to get rid of the scalar field. For the static case, the composition of the $SO(2)$-transformation (\ref{eq:sigma_so2_transformation}) and the $EG$-transformations applied to the vacuum solution give another vacuum solution with an appropriate choice of the transformation parameters. Then such transformation is the so-called Zipoy-Voorhees transformation $\psi \to s\psi$, $\gamma^\psi \to s^2 \gamma^\psi$.

Moreover, the equation of motion of the scalar field does not contain explicitly the coordinate $z$: 
\begin{equation}\label{eq:eom_weyl_phi}
    \partial_\rho \rho \partial\rho \phi + \rho \partial_z^2 \phi = 0.
\end{equation}
So, if $\phi_0$ is a solution, then $\phi_n = \partial_z^n \phi$ is a solution too. Regarding to the problem of Kerr-like solutions with scalar fields, such transformations was considered in \cite{Chauvineau:2018zjy}.

For example, the scalar field (\ref{eq:rfzv_solution_phi}) in the Weyl coordinates has a complicated form
\begin{subequations}
\begin{align}\label{eq:rfzv_solution_phi_weyl}
    &
    \phi(r) = \phi_\infty + \frac{\Sigma }{2 b} \log
    \frac{ z + \text{sign}(z)\,\Phi(\rho,z) }{ z - \text{sign}(z)\,\Phi(\rho,z) },
    \\ & 
    \Phi(\rho,z) = \frac{1}{\sqrt{2}}\sqrt{b^2+\rho ^2+z^2 - \sqrt{2 \rho ^2 \left(b^2+z^2\right)+\left(b^2-z^2\right)^2+\rho ^4}}.
\end{align}
\end{subequations}
Acting with $\partial_z = r_{,z} \partial_r + \theta_{,z} \partial_\theta $ on (\ref{eq:rfzv_solution_phi_weyl}) and using the coordinate transformations $\rho=\sqrt{\Delta}\sin\theta$, $z=(r-M)\cos\theta$, after lengthy calculations one can obtain
\begin{subequations}\label{eq:rfzv_solution_1}
\begin{align}
    &\label{eq:rfzv_solution_phi_1}
    \phi_{(1)} = \frac{\Lambda_{(1)} \cos\theta}{\Delta + b^2 \sin^2\theta},\\
    &\label{eq:rfzv_solution_gamma_1}
    \gamma^\phi_{(1)} =\frac{-\Lambda_{(1)}^2}{8 b^4\left(\Delta + b^2 \sin^2\theta\right)^2}\left[
    \Delta^2 + 2 b^2\Delta \sin^2\theta - b^4 \sin^4\theta
    +\frac{
        4 b^4 \sin ^2(2 \theta ) (r-M)^2 \Delta 
    }{
        \left(\Delta + b^2\sin^2\theta\right)^2
    }\right],
\end{align}
\end{subequations}
where $\Lambda_{(1)}$ is a constant. The asymptotic behaviour of the scalar field $\phi \approx \Lambda_{(1)} \cos\theta/r^2$ suggests that the solution describes the rotating source with a scalar dipole moment. 

Applying the EG-duality to the Kerr metric gives the solutions (\ref{eq:rfzv_solution}), (\ref{eq:rfzv_chauvineau}) or (\ref{eq:rfzv_solution_1}) if we choose $\phi$ in the form (\ref{eq:rfzv_solution_phi}), (\ref{eq:rfzv_chauvineau_phi}) or (\ref{eq:rfzv_solution_phi_1}) correspondingly. Scalar field satisfies the Laplace equation, thus we can represent the solution in terms of the rod structure. The rod structure of rotating generalization of FJNW solution is not clear for the moment. Note that we are not restricted to use $\phi$ depending on the coordinate $r$ only, but this case seems to be the most relevant.

As the result of the $SO(2)$-symmetry (\ref{eq:sigma_so2_transformation}), in the static case $\omega=0$, the equations of motion for $\gamma^\psi$ and $\psi$ have the  form (\ref{eq:eom_weyl_gamma_phi}), (\ref{eq:eom_weyl_phi}) similar to $\gamma^\phi$ and $\phi$:
\begin{subequations}
    \begin{equation}\label{eq:eom_weyl_gamma_psi}
        \gamma^\psi_{,\rho} =\rho\left( \psi_{,\rho}^2 - \psi_{,z}^2 \right),\qquad
        \gamma^\psi_{,z} =2\rho \psi_{,\rho}\psi_{,z},
    \end{equation}
    \begin{equation}\label{eq:eom_weyl_psi}
        \partial_\rho \rho \partial\rho \psi + \rho \partial_z^2 \psi = 0.
    \end{equation}
\end{subequations}
Let us consider a stationary generalization of ZV solution with an arbitrary   $\delta$, an angular momentum $J$ and some potentials $\psi$, $\gamma^\psi$, $\omega$. We will split $\gamma^\psi$ and $\psi$ into two parts: the static limit $\gamma^s$, $\psi^s$ and the rotational part $\gamma^\omega$, $\psi^\omega$
\begin{equation*}
    \gamma^s = \lim_{J\to 0} \gamma^\psi = \delta^2 \gamma_{Sch},\qquad
    \psi^s = \lim_{J\to 0} \psi = \delta \psi_{Sch},\qquad
    \gamma^\omega = \gamma^\psi - \gamma^s,\qquad
    \psi^\omega = \psi - \psi^s,
\end{equation*}
where $\gamma_{Sch} = \frac{1}{2}\ln\frac{L^2-k^2}{l_{+}l_{-}}$, $\psi_{Sch} = \frac{1}{2}\ln\frac{L-k}{L+k}$ are the potentials of the Schwarzschild solution (note that $\gamma_{Sch}$ corresponds to FJNW solution as well), $L=\frac{1}{2}(l_{+}+l_{-})$, $l_{\pm}=\sqrt{\rho^2 + (z\pm k)^2}$, with $k$ being a constant entering the spheroidal coordinates. Since $\psi^s, \gamma^s$ and $\phi$, $\gamma^\phi$ satisfy the same equations, we can introduce a scalar field into our solution in the form
\begin{equation*}
    \gamma^\phi = c^2 \gamma_{Sch},\qquad
    \phi = c \psi_{Sch}.
\end{equation*}
The final metric is described by functions $\psi = \delta \psi_{Sch} + \psi^\omega$, $\phi = c \psi_{Sch}$, $\omega$, $\gamma = (\delta^2 + c^2)\psi_{Sch} + \gamma^\omega$. Setting $\delta^2 + c^2 = 1$, we get rid of the 3-metric deformation of   ZV kind. On   one hand, in the static limit $\omega, \psi^\omega, \gamma^\omega \to 0$ such a solution exactly corresponds to FJNW (this was noticed by Eris and Gurses \cite{Eris:1976xj}). For zero scalar case $c=0$, the constraint gives $\delta=1$ and the solution represent the rotating Schwarzschild solution, i.e. the Kerr metric. Thus, such a solution can be considered as a full-fledged rotating generalization of FJNW. On the other hand, one can take the scalar field in the form $\phi = c \psi_{Sch} + Q(J) \tilde{\phi}$, where $Q(0)=0$ is some function of the angular momentum $J$, and $\tilde{\phi}$ is an arbitrary solution of (\ref{eq:eom_weyl_phi}). Such a solution will have the same properties for static and scalarless limits. Here we will not consider this case.

Rotating ZV solutions were reported also by Hori in \cite{Hori:1996} without any evidence of their correctness. Eris and Gurses mentioned without details the possibility to apply their transformation to Tomimatsu-Sato solutions. Tomimatsu Sato themselves presented   rotating vacuum generalization of ZV with integer deformation parameter $\delta=1,2,3,4$ \cite{Tomimatsu:73} and Hori generalized these solutions for arbitrary integer $\delta \in \mathbb{Z}_+$  \cite{Hori:1977js}. From the constraint, it follows that $c^2 < 0$ for ${\delta > 1}$, and we can construct the rotating FJNW from the Tomimatsu-Sato solutions for a phantom scalar field only. We will consider the case $\delta = 2$ with
\begin{equation}
    e^{2\psi} = \frac{A}{B},\qquad
    e^{2\gamma^\psi} = \frac{A}{p^4(x^2-y^2)^4},\qquad
    e^{2\gamma^\phi} = \left(\frac{x^2-y^2}{x^2-1}\right)^3,\qquad
    \omega = \frac{2q M (1-y^2)C}{A},
\end{equation}
where the constants satisfy the constraint $p^2+q^2=1$ and the functions $A$, $B$, $C$ are
\begin{align}
    A = & p^4 (x^2-1)^4 + q^4(1-y^2)^4 - 
    \\\nonumber &
    - 2p^2q^2(x^2-1)(1-y^2)\left[
        2(x^2-1)^2 + 2(1-y^2)^2 + 3(x^2-1)(1-y^2)
    \right],
    \\
    B = & \left[
        p^2(x^2+1)(x^2-1)-q^2(y^2+1)(1-y^2)+2px(x^2-1)
    \right]^2 +
    \\\nonumber &
    + 4q^2y^2\left[
        px(x^2-1) + (px+1)(1-y^2)
    \right]^2,
    \\
    C = & -p^3x(x^2-1)\left[
        2(x^2+1)(x^2-1) + (x^2+3)(1-y^2)
    \right] -
    \\\nonumber &
    -p^2(x^2-1)\left[
        4x^2(x^2-1) + (3x^2+1)(1-y^2)
    \right]
    +q^2(px+1)(1-y^2)^3.
\end{align}
The final solution reads:
\begin{align}\label{eq:rF}
    &ds^2 = -\frac{A}{B}\left(dt-\omega d\varphi\right)^2+\frac{B}{A} k^2 \left(
        H (dx^2 + h^{-1}dy^2) + (x^2-1)(1-y^2)d\varphi^2
    \right),
    \\\nonumber&
    H = \left(1 + \frac{q^2}{p^2}h^2\right)^2 - 4\frac{q^2}{p^2}h(h+1)^2,
    \\\nonumber&
    \phi = \pm \frac{i}{2} \sqrt{3} \ln \frac{x-1}{x+1},\qquad
    h = \frac{1-y^2}{x^2-1},\qquad
    k = M p/2,\qquad
    q=J/M^2,
\end{align}
where $J=Ma$ is the angular momentum and $M$ is the mass, $a$  is the rotation parameter. Note that $k^2 H = e^{2\gamma}(x^2-y^2)/(x^2-1)$ due to coordinate transformation from $\rho, z$ to $x, y$. From the definition of the scalar charge $\phi \approx \phi_\infty + \Sigma / r$ at infinity, one can find  $\Sigma = \mp i \sqrt{3}Mp$.

The solution (\ref{eq:rF}) has a horizon at $x = \pm 1$, an ergo-region is defined by the equation $H=0$, which can be resolved as $h=\tilde{h}(q^2/p^2)$, and singularity:
\begin{equation} \label{eq:r_ts}
    R = 2g^{xx}(\partial_x \phi)^2 = \frac{-6A}{B H k^2(x^2-1)^2} = \frac{-24p^2}{M^2} \frac{ (x^2-1)^2 }{ B }.  
\end{equation}
From inspection of the Ricci scalar, the solution has singularities at $B=0$, while $B$ is a sum of two squares. Therefore, both terms should be equal to zero
\begin{subequations}
    \begin{equation} \label{eq:r_ts_cond_1}
        p^2(x^2+1)(x^2-1)-q^2(y^2+1)(1-y^2)+2px(x^2-1) = 0,
    \end{equation}
    \begin{equation} \label{eq:r_ts_cond_2}
    y\left( px(x^2-1) + (px+1)(1-y^2) \right) = 0.
    \end{equation}
\end{subequations}
The second equation (\ref{eq:r_ts_cond_1}) holds if $y=0$ or $y^2 = (1 + p x^3)/(1 + p x)$. Let us start from the second solution, substituting $y^2$ into the first equation (\ref{eq:r_ts_cond_1})
\begin{equation*}
    \frac{p \left(x^2-1\right) \left(4 p^2 x^3+p \left(x^4+6 x^2+1\right)+4 x\right)}{(p x+1)^2} = 0.
\end{equation*}
The first root $x\pm1$ does not lead to divergence due to  presence of $(x^2-1)^2$ in the numerator of Ricci scalar (\ref{eq:r_ts}). The second bracket is positive for $p>0$, $x > 0$. Therefore, this case can create a singularity under the horizon $x=1$ only.
The first solution $y=0$ substituted into the first equation (\ref{eq:r_ts_cond_1}) gives a condition
\begin{equation*}
    p^2 x^4 + 2 p x (x^2-1) - 1 = 0
\end{equation*}
which has one root in the outer region $x>1$.
Thus, the solution represents a regular black hole except for a singular ring on the equator. The scalar field diverges on the horizon, so the horizon has a scalar charge, which is typical for the static FJNW solution with $S>1$. Moreover, the scalar field is regular in the ring singularity, so the ring does not carry the scalar charge.

The metric functions depend on $p^2, q^2, x^2, y^2$, $px$ and allow for analytical continuation $x\to i x$, $p\to i p$, (i.e. $a>M$), under which the metric remains physical with the same signature, but the scalar field becomes real $\phi = \pm\sqrt{3}\text{ arctan } x$ (up to the additive constant). In this case, $x\geq0$ and $-1\leq y \leq +1$ represent the oblate spheroidal coordinate system, and the scalar field has a cusp at the disk $x=0$. Therefore, this solution cannot be considered as the rotating Fisher generalization. We can analytically continue the coordinate $x$ for the whole real numbers $x\in\mathbb{R}$ to get a wormhole without scalar cusp. Such objects with wormhole interpretation were described by Gibbons and Volkov in \cite{Gibbons:2016bok}.

\section{Conclusions} \label{sec:conclusions}
Let us summarize briefly our results. First, we obtained a new static generalization of FJNW endowed with the oblateness and NUT parameters using the most transparent sigma-model formulation of the field equations. 
Joined ZV-FJNW solution opened a way to apply Cl\'ement's technique for generating rotation, obtaining non-trivial result without Maxwell field. In this solution the oblateness parameter can not set to unity, because of the intrinsic constraint of the CT. However the solution is simple and can be used as the legitimate solution for studying physics beyond the Kerr paradigm. In the extremal rotation limit our solution coincides with one of solutions that found recently by Chauvineau.

As an independent approach, we used hidden symmetry of stationary axisymmetric solution rediscovered by Eris and Gurses. Applying the EG duality, we could reproduce the result obtained via CT. Using EG-transformation, we put forward the argument that the rotating ZV solutions are dual to rotating FJNW solutions. As an example, we obtained rotating FJNW solution dual to Tomimatsu-Sato solution with $\delta=2$. Such solution represented a regular black hole with a ring singularity around the horizon and phantom scalar field. Using the complex transformations, the scalar field can be turned into real field, but the solution is not a generlization of the FJNW solution anymore and it can be interpreted as a disk with scalar charge or a wormhole with ring singularity.

Also, we obtained one more solution with generating technique suggested by Chauvineau for Kerr-like metrics, which can be enlarged for any axisymmetric solution using the EG-transformation. This technique produces a new solution for the scalar field and allows to find exact correction to the metric in terms of integrals. We applied this technique to the solution obtained with Cl\'ement's technique.

All solutions we obtained in this paper can be considered as a correct alternative to those discussed in Appendicies.

\noindent {\textbf{\textit {Acknowledgments.}}}
The authors are grateful to G\'erard Cl\'ement for careful reading of the manuscript and valuable comments. The work was supported by the Russian Foundation for Basic Research on the project 17-02-01299a. The authors would like to acknowledge networking support by the COST Action CA16104.

\appendix
\setcounter{equation}{0}
\section{False rotating FJWN}\label{sec:false rotating_fjnw}
Consider the metric, obtained in  \cite{Agnese:1985xj} by application of the JN trick to FJNW. In terms of the sigma-model variables (\ref{eq:sigma_ansatz}) it reads
\begin{subequations}\label{eq:agnese_metric}
\begin{equation}
    f = \frac{R^2}{\sigma^2} \left(
        \Delta - \tilde{\omega}^2 \sin^2\theta
    \right),\qquad
    \omega_i dx^i = -f^{-1} \tilde{\omega} \sin^2\theta d\varphi,
\end{equation}
\begin{equation}
    h_{ij}dx^i dx^j = \frac{fR^2}{\Delta} \left( dr^2 + \Delta d\theta^2\right) + \Delta \sin^2\theta d\varphi^2,
\end{equation}
\end{subequations}
with the scalar field
\begin{equation}\label{eq:agnese_phi}
    \phi(r) = \frac{\Sigma}{2\sqrt{\eta^2 - a^2}}\ln \left( 1 - \frac{\eta + \sqrt{\eta^2 - a^2}}{r} \right), \qquad
    \eta = \sqrt{M^2 + \Sigma^2},
\end{equation}
where
\begin{equation*}
    \tilde{\omega} = \frac{a(R^2 + a^2\sin^2\theta - \Delta )}{R^2},\qquad
    R^2 = (r^2 + a^2\cos^2\theta)\left(
        1 - \frac{2\eta r}{r^2 + a^2 \cos^2\theta}
    \right)^{1-M/\eta},
\end{equation*}
\begin{equation*}
    \Delta = r^2 - 2\eta r + a^2,\qquad
    \sigma^2 = (R^2 + a^2 \sin^2\theta)^2 - \Delta a^2 \sin^2\theta.
\end{equation*}
For $\Sigma=0$, $\eta = M$, and we recover the Kerr  solution. This metric looks simple and it became a popular model for describing possible deviations from General Relativity in astrophysical observations \cite{Gyulchev:2008ff,Kovacs:2010xm,Liu:2018bfx,Jusufi:2018gnz}. Pirogov \cite{Pirogov:2013wia} verified part of the Einstein equations and found that they are not satisfied. This claim was supported in   \cite{Hansen:2013owa}. Here we check the sigma-model equations.
Considering the equation $\Box \phi = 0$ for $\phi$ depending on $r$ only, taking into account $\sqrt{-g} = R^2 \sin\theta$, one can derive the equation
\begin{equation}
    \partial_r \left(\Delta \partial_r \phi \right) = 0,
\end{equation}
which can be solved with $\phi = \text{const} \ln(r - \eta + \sqrt{\eta^2 - a^2})/(r - \eta - \sqrt{\eta^2 - a^2})$, but not (\ref{eq:agnese_phi}). Still this does not  mean  that the metric (\ref{eq:agnese_metric}) is incorrect.

The $\sigma$-model (\ref{eq:sigma_metric}) implies the following equation   for $\psi$
\begin{equation}
    \Delta \psi + \frac{1}{2} e^{-4\psi} (\partial \chi)^2 = 0, 
\end{equation}
where $\Delta$ and the contraction  over indices relate to the 3-metric:
\begin{equation}
    (\partial\chi)^2 = e^{8\psi} \left( (\partial_i \omega_j)(\partial^i \omega^j) - (\partial_i \omega_j)(\partial^j \omega^i)\right),
\end{equation}
for $\omega_i dx^i = \omega(r,\theta) d\varphi$ and diagonal 3-metric the second term $(\partial_i \omega_j)(\partial^j \omega^i)$ is zero. The first term we will write as $(\partial \omega)^2$. Then the equation is
\begin{equation}
    \Delta \psi + \frac{1}{2} e^{4\psi} (\partial \omega)^2 = 0.
\end{equation}
It can be expanded as
\begin{equation}
    \partial_r \left( \Delta \partial_r \psi \right)
    + \frac{1}{\sin\theta}\partial_\theta \left( \sin\theta \partial_\theta \psi \right)
    + \frac{e^{4\psi}}{2\sin^2\theta}\left(
        (\partial_r \omega)^2
        + (\partial_r \omega)^2/\Delta
    \right) = 0
\end{equation}
Substituting the functions $f$ and $\omega$ and expanding as $r\to infty$ we find the non-zero term
\begin{equation}
    \frac{a^2 M (3 \cos (2 \theta )+5) (M-\eta )}{r^4} + \mathcal{O}(r^{-5})=0.
\end{equation}
  This can be fulfilled for $a=0$ (static FJNW solution) or $\eta=M$ (Kerr). So we confirm the results of \cite{Pirogov:2013wia} and \cite{Hansen:2013owa}.

\section{False rotating FJNW in Brans-Dicke}
Another rotating solution with the scalar field generated with JN algorithm was derived within the Brans-Dicke theory \cite{Krori:1982}. The Brans-Dicke equations of motion read
\begin{subequations}
\begin{equation}\label{eq:bd_eom_phi}
    \Box\Phi = 0,
\end{equation}
\begin{equation}\label{eq:bd_eom_r}
    R_{\mu\nu} - \frac{1}{2}g_{\mu\nu}R =
    \frac{\omega}{\Phi^2} \left(
        \Phi_{;\mu}\Phi_{;\nu}
        - \frac{1}{2}g_{\mu\nu} \Phi_{;\lambda}\Phi^{;\lambda}
    \right)
    + \frac{1}{\Phi} \left(
        \Phi_{;\mu\nu} - g_{\mu\nu}\Box\Phi
    \right).
\end{equation}
\end{subequations}
Taking into account (\ref{eq:bd_eom_phi}), one can find Ricci tensor
\begin{equation}
    R_{\mu\nu} = \frac{\omega}{\Phi^2} \partial_\mu \Phi \partial_\nu \Phi + \frac{1}{\Phi} \Phi_{;\mu\nu}.
\end{equation}
The theory can be formulated in the Jordan and Einstein frames,the  Einstein frame  corresponding to MES. To change the frame, one has to make the transformations
\begin{equation}
    g^{E}_{\mu\nu} = \Phi g^{J}_{\mu\nu},\qquad
    \phi = \frac{1}{2}\sqrt{2\omega + 3} \ln \Phi.
\end{equation}
  The FJNW solution in the Jordan frame reads
\begin{equation}\label{eq:fisher_jordan}
    ds^2 = -F^{-\sigma+S} dt^2 + F^{-\sigma-S}\left(dr^2 + r^2 F \left(d\theta^2 + \sin^2\theta d\varphi^2\right)\right),\qquad
    \Phi = \Phi_0 F^\sigma
\end{equation}
\begin{equation}
    F=1-\frac{2M}{Sr},\qquad
    S = \frac{M}{\sqrt{M^2 + \Sigma^2}},\qquad
    \sigma = -\frac{\Sigma S}{ M \sqrt{2\omega+3}}.
\end{equation}

The solution found by Krori and Bhattacharjee in \cite{Krori:1982} by application of the JN trick is
\begin{subequations}\label{eq:krori_solution}
\begin{equation} \label{eq:krori_metric}
    ds^2 = f_K^\eta (dt - \omega_K d\varphi)^2 - f_K^\xi \rho (dr^2/\Delta + d\theta^2 +\sin^2\theta d\varphi^2) + 2 f_K^\sigma \omega_K (dt - \omega_K d\varphi) d\varphi,
\end{equation}
\begin{equation}\label{eq:krori_field}
    \Phi = \Phi_0 f_K^\sigma,
\end{equation}
\end{subequations}
where
\begin{equation}
    f_K = 1 - 2r_0 r / \rho,\qquad
    \rho = r^2 + a^2 \cos^2\theta,\qquad
    \omega_K = a \sin^2\theta,\qquad
    \Delta = r(r-2r_0) + a^2
\end{equation}
and
\begin{equation}
    \sigma = (\eta + \xi - 1)/2 = -c/2\lambda,\qquad
    \eta = 1/\lambda,\qquad
    \xi = (\lambda - c - 1)/\lambda
\end{equation}
with free parameters $\lambda$ and $c$. The static limit of the solution (\ref{eq:krori_field}) should coincide with (\ref{eq:fisher_jordan}) up to the definition of constants. The solution (\ref{eq:fisher_jordan}) possesses a property that $\ln|g_{tt}g_{rr}| = - 2 \ln \Phi + \text{const}$. For the solution (\ref{eq:krori_solution}) we find 
\begin{equation*}
    \ln|g_{tt}g_{rr}| = (\eta + \xi - 1) \ln f_K = 2\ln\Phi + \text{const},
\end{equation*}
thus the scalar field is incorrect and the correct one is $\Phi = \Phi_0 f_K^{-\sigma}$.

For $tt$-component, the Einstein equation is
\begin{equation}\label{eq:bd_eq_tt}
    R_{tt} = \frac{1}{\Phi}\Phi_{;tt} = -\Gamma^{r}_{tt} \partial_r \ln \Phi.
\end{equation}
Let us calculate an asymptotic behavior of the quantity $X = R_{tt} - \Phi_{;tt}/\Phi$ up to the 7th order for the solution (\ref{eq:krori_solution}). The first non-zero term of the Taylor series starts from the 4th order. The 4th and the 5th terms are zero if we use a corrected definition of the scalar field. Then the term of the 6th order reads
\begin{equation}
    X \approx \frac{a^2 r_0^2 \left((c+2)^2-4 \lambda \right) (5 + 3 \cos2\theta)}{4 \lambda ^2 r^6} + \mathcal{O}(r^{-7}),
\end{equation}
which is zero for $a=0$ or $\lambda = (c+2)^2/4$. The first case brings us back to the static solution and satisfies the equation (\ref{eq:bd_eq_tt}) exactly. Substituting the second case into the 7th order gives
\begin{equation}
    X \approx \frac{64 a^2 c r_0^3 \cos^2\theta}{(c+2)^3 r^7} + \mathcal{O}(r^{-8})
\end{equation}
and requires either $c$ or $r_0$ to be zero, which guarantees the trivial form of the scalar field. Therefore, the solution found with JN algorithm in \cite{Krori:1982} is incorrect as well.



\begin{thebibliography}{9}
\bibitem{Galtsov:2018xuc}
  D.~Gal'tsov and S.~Zhidkova,
  {\it Ghost-free Palatini derivative scalar-tensor theory: Desingularization and the speed test},
  Phys.\ Lett.\ B {\bf 790} (2019) 453,
  [1808.00492].

\bibitem{BenAchour:2019fdf} 
  J.~Ben Achour, H.~Liu and S.~Mukohyama,
  {\it Hairy black holes in DHOST theories: Exploring disformal transformation as a solution-generating method},
  [1910.11017].
  
\bibitem{Fisher:1948yn}
  I.~Z. Fisher, 
  { \it Scalar mesostatic field with regard for gravitationa  effects},
  Zh.~Eksp.~Teor.~Fiz. {\bf 18} (1948) 636, [gr-qc/9911008].

\bibitem{Bergmann:1957zza} 
  O.~Bergmann and R.~Leipnik,
  {\it Space-Time Structure of a Static Spherically Symmetric Scalar Field},
  Phys.\ Rev.\  {\bf 107} (1957) 1157.
  
\bibitem{Penney:1968zz} 
  R.~Penney,
  { \it Axially Symmetric Zero-Mass Meson Solutions of Einstein Equations},
  Phys.\ Rev.\  {\bf 174} (1968) 1578.
  
  \bibitem{Janis:1968zz} 
  A.~I.~Janis, E.~T.~Newman and J.~Winicour,
  {\it Reality of the Schwarzschild Singularity},
  Phys.\ Rev.\ Lett.\  {\bf 20} (1968) 878.

\bibitem{Penney:1969xk} 
  R.~Penney,
  {\it Generalization of the Reissner-Nordstroem solution to the Einstein field equations},
  Phys.\ Rev.\  {\bf 182} (1969) 1383.
  
\bibitem{Deser:1969nv} 
  S.~Deser and J.~Higbie,
  {\it Essential singularities in general relativity},
  Phys.\ Rev.\ Lett.\  {\bf 23} (1969) 1184.
  
\bibitem{Wyman:1981bd}
 M.~Wyman, 
 {\it Static Spherically Symmetric Scalar Fields in General Relativity}, 
 Phys.~Rev.~D {\bf 24} (1981) 839.
 
\bibitem{Virbhadra:1997ie} 
  K.~S.~Virbhadra,
  {\it Janis-Newman-Winicour and Wyman solutions are the same},
  Int.\ J.\ Mod.\ Phys.\ A {\bf 12} (1997) 4831,
  [gr-qc/9701021].

\bibitem{Bhadra:2001fx} 
  A.~Bhadra and K.~K.~Nandi,
  {\it On the equivalence of the Buchdahl and the Janis-Newman-Winicour solutions},
  Int.\ J.\ Mod.\ Phys.\ A {\bf 16} (2001) 4543.
  
\bibitem{Abdolrahimi:2009dc} 
  S.~Abdolrahimi and A.~A.~Shoom,
  {\it Analysis of the Fisher solution},
  Phys.\ Rev.\ D {\bf 81} (2010) 024035,
  [0911.5380].

\bibitem{Faraoni:2015paa} 
  V.~Faraoni, A.~Prain and A.~F.~Zambrano Moreno,
  {\it Black holes and wormholes subject to conformal mappings},
  Phys.\ Rev.\ D {\bf 93}, 2 (2016) 024005,
  [1509.04129].
  
\bibitem{Domenech:2019syf} 
  G.~Dom\`enech, A.~Naruko, M.~Sasaki and C.~Wetterich,
  {\it Could the black hole singularity be a field singularity?},
  [1912.02845].

\bibitem{Agnese:1985xj} 
  A.~G.~Agnese and M.~La Camera,
  {\it Gravitation Without Black Holes},
  Phys.\ Rev.\ D {\bf 31} (1985) 1280.
 
\bibitem{Newman:1965tw} 
  E.~T.~Newman and A.~I.~Janis,
  {\it Note on the Kerr spinning particle metric},
  J.\ Math.\ Phys.\  {\bf 6} (1965) 915.
  
\bibitem{Erbin:2016lzq} 
  H.~Erbin,
  {\it Janis-Newman algorithm: generating rotating and NUT charged black holes},
  Universe {\bf 3}, 1 (2017) 19,
  [1701.00037].

\bibitem{Gyulchev:2008ff}
  G.~N. Gyulchev and S.~S. Yazadjiev, 
  {\it Gravitational Lensing by Rotating Naked Singularities}, 
  Phys.~Rev.~D {\bf 78} (2008) 083004,
  [0806.3289].
  
\bibitem{Kovacs:2010xm}
  Z.~Kovacs and T.~Harko, 
  {\it Can accretion disk properties observationally  distinguish black holes from naked singularities},
  Phys.~Rev.~D {\bf 82} (2010) 124047,
  [1011.4127].

\bibitem{Liu:2018bfx} 
  H.~Liu, M.~Zhou and C.~Bambi,
  {\it Distinguishing black holes and naked singularities with iron line spectroscopy},
  JCAP {\bf 1808}, 08 (2018) 044,
  [1801.00867].

\bibitem{Jusufi:2018gnz} 
  K.~Jusufi, A.~Banerjee, G.~Gyulchev and M.~Amir,
  {\it Distinguishing rotating naked singularities from Kerr-like wormholes by their deflection angles of massive particles},
  Eur.\ Phys.\ J.\ C {\bf 79}, 1 (2019) 28,
  [1808.02751].
  
\bibitem{Pirogov:2013wia} 
  Y.~F.~Pirogov,
  {\it Towards the rotating scalar-vacuum black holes},
  [1306.4866].
  
\bibitem{Hansen:2013owa} 
  D.~Hansen and N.~Yunes,
  {\it Applicability of the Newman-Janis Algorithm to Black Hole Solutions of Modified Gravity Theories},
  Phys.\ Rev.\ D {\bf 88}, 10 (2013) 104020,
  [1308.6631].

\bibitem{Chauvineau:2018zjy} 
  B.~Chauvineau,
  { \it New method to generate exact scalar-tensor solutions},
  Phys.\ Rev.\ D {\bf 100}, 2 (2019) 024051,
  [1812.04934].
 
\bibitem{Tiwari:1976vc} 
  R.~N.~Tiwari and B.~K.~Nayak,
  {\it Class of the Brans-Dicke Maxwell Fields},
  Phys.\ Rev.\ D {\bf 14} (1976) 2502.
  
\bibitem{Kim:1998hc} 
  H.~Kim,
  {\it New black hole solutions in Brans-Dicke theory of gravity},
  Phys.\ Rev.\ D {\bf 60} (1999) 024001,
  [gr-qc/9811012].
  
\bibitem{Kirezli:2015wda} 
  P.~Kirezli and O.~Delice,
  {\it Stationary axially symmetric solutions in Brans-Dicke theory},
  Phys.\ Rev.\ D {\bf 92} (2015) 104045,
  [1507.00910].
  
\bibitem{Krori:1982} 
  K.~D.~Krori and D.~R.~Bhattacharjee,
  {\it Kerr‐like metric in Brans–Dicke theory},
  J. Math. Phys. {\bf 23} (1982) 637.

\bibitem{Janis:1970kn} 
  A.~I.~Janis, D.~C.~Robinson and J.~Winicour,
  {\it Comments on Einstein scalar solutions},
  Phys.\ Rev.\  {\bf 186} (1969) 1729.

\bibitem{Buchdahl:1959nk} 
  H.~A.~Buchdahl,
  {\it Reciprocal Static Metrics and Scalar Fields in the General Theory of Relativity},
  Phys.\ Rev.\  {\bf 115} (1959) 1325.
  
\bibitem{Galtsov:1995mb} 
  D.~V.~Galtsov, A.~A.~Garcia and O.~V.~Kechkin,
  {\it Symmetries of the stationary Einstein-Maxwell dilaton theory},
  Class.\ Quant.\ Grav.\  {\bf 12} (1995) 2887,
  [hep-th/9504155].
  
\bibitem{Clement:1997tx} 
  G.~Cl\'ement,
  {\it From Schwarzschild to Kerr: Generating spinning Einstein-Maxwell fields from static fields},
  Phys.\ Rev.\ D {\bf 57} (1998) 4885,
  [gr-qc/9710109].

\bibitem{Zipoy:66}
  D.~M. Zipoy, 
  {\it Topology of Some Spheroidal Metrics},
  J. Math. Phys. {\bf 7} (1966) 1137.
 
\bibitem{Voorhees:70}
  B.~H. Voorhees, 
  {\it Static Axially Symmetric Gravitational Fields},
  Phys. Rev. D {\bf 2} (1970) 2119.

\bibitem{Xanthopoulos:1989kb} 
  B.~C.~Xanthopoulos and T.~Zannias,
  {\it Einstein Gravity Coupled to a Massless Scalar Field in Arbitrary Space-time Dimensions},
  Phys.\ Rev.\ D {\bf 40} (1989) 2564.

\bibitem{Garfinkle:1990qj} 
  D.~Garfinkle, G.~T.~Horowitz and A.~Strominger,
  {\it Charged black holes in string theory},
  Phys.\ Rev.\ D {\bf 43} (1991) 3140;
  {\bf 45} (1992) 3888 [erratum].

\bibitem{Winicour:1968zz} 
  J.~Winicour, A.~I.~Janis and E.~T.~Newman,
  {\it Static, Axially Symmetric Point Horizons},
  Phys.\ Rev.\  {\bf 176} (1968) 1507.

\bibitem{Eris:1976xj} 
  A.~Eris and M.~Gurses,
  {\it Stationary Axially Symmetric Solutions of Einstein-Maxwell Massless Scalar Field Equations},
  J.\ Math.\ Phys.\  {\bf 18} (1977) 1303.

\bibitem{Tomimatsu:73}
  A.~Tomimatsu and H.~Sato,
  {\it New Series of Exact Solutions for Gravitational Fields of Spinning Masses},
  Prog. Theor. Phys., {\bf 50}, 1 (1973) 95-110.

\bibitem{Ehlers:1961zza} 
  J.~Ehlers,
  {\it Transformations of static exterior solutions of Einstein's gravitational field equations into different solutions by means of conformal mapping},
  Colloq.\ Int.\ CNRS {\bf 91} (1962) 275.

\bibitem{Stelea:2006zz} 
  C.~I.~Stelea, {\it Higher dimensional Taub-NUT spaces and applications}, PhD Thesis,  University of Waterloo, 2006.

\bibitem{Kobyalko:2017zz} 
  K.~V.~Kobyalko, {\it Black holes beyond the Standard Model}, Master Thesis, Lomonosov Moscow State University, 2017 (in Russian).

\bibitem{Chazy:1924} 
  J.~Chazy, {\it Sur la champ de gravitation de deux masses fixes dans la th\'eory de la relativit\'e},
  Bull. Soc. Math. France {\bf 52} (1924) 17.

\bibitem{Curzon:1925} 
  H.~Curzon, {\it Cylindrical solutions of Einstein's gravitation equations},
  P. Lond. Math. Soc. {\bf 23} (1925) 477–480.

\bibitem{Neugebauer:69}
  G.~Neugebauer and D.~Kramer,
  {\it Eine Methode zur Konstruktion stationärer Einstein‐Maxwell‐Felder},
  Ann. der Physik (Leipzig), {\bf 24} (1969) 62.

\bibitem{Hori:1996}
  Shoichi Hori, 
  {\it Generalization of Tomimatsu-Sato Solutions},
  Prog. Theor. Phys., {\bf 95}, 1 (1996) 65-70.

\bibitem{Hori:1977js} 
  S.~Hori,
  {\it On the Exact Solution of Tomimatsu-Sato Family for an Arbitrary Integral Value of the Deformation Parameter},
  Prog.\ Theor.\ Phys.\  {\bf 59} (1978) 1870;
  {\bf 61} (1979) 365 [erratum].

\bibitem{Gibbons:2016bok} 
  G.~W.~Gibbons and M.~S.~Volkov,
  {\it Ring wormholes via duality rotations},
  Phys.\ Lett.\ B {\bf 760} (2016) 324,
  [1606.04879].

\end{thebibliography}
\end{document}